\documentclass[12pt,a4paper]{JHEP3}
\usepackage{amsmath,amssymb,bm}
\usepackage{latexsym}
\usepackage{graphicx}

\def\eqref#1{Eq.~(\ref{#1})}

\def\Eq#1{\begin{equation} #1 \end{equation}}
\def\Eqr#1{\begin{eqnarray} #1 \end{eqnarray}}
\def\Eqrsubl#1#2{\begin{subequations}\label{#1}\Eqr{#2}\end{subequations}}

\newcommand{\nn}{\nonumber}
\newcommand{\pd}{\partial}

\def\Ysp{{\rm Y}}

\def\X5sp{{\rm X}_5}
\def\Y3sp{{\rm Y}_3}
\def\Z3sp{{\rm Z}_3}

\def\e{{\rm e}}

\newcommand{\bea}{\begin{eqnarray}}
\newcommand{\eea}{\end{eqnarray}}
\newcommand{\beq}{\begin{equation}}
\newcommand{\eeq}{\end{equation}}

\title{
No-Go theorems for ekpyrosis from ten-dimensional supergravity
}

\author{
Kunihito Uzawa\\
Department of Physics,
School of Science and Technology,\\
~~Kwansei Gakuin University, Sanda, Hyogo 669-1337, Japan.}

\abstract{%
 In this note we investigate whether the new ekpyrotic 
scenario can be embedded into ten-dimensional supergravity. 
We use that the scalar potential obtained from flux 
compactifications of type II supergravity with sources 
has a universal scaling with respect to the dilaton and 
the volume mode. Similar to the investigation of 
inflationary models, we find very strong constraints 
ruling out ekpyrosis from analysing the fast-roll 
conditions. We conclude that flux compactifications tend 
to provide potentials that are neither too flat and positive 
(inflation) nor too steep and negative (ekpyrosis). 
}

\begin{document}


\section{Introduction}
 \label{sec:introduction}
The strong no-go theorems which exclude tree-level 
de Sitter compactifications under a few simple assumptions  
with or without negative tension objects such as orientifold planes 
have been much explored because of the possible 
cosmological and phenomenological interests. 
However, the no-go theorems for ekpyrotic scenario which is 
alternative to inflation model in string theory 
is much less extensive. 
One motivation for the present work is to improve this situation. 
Since orientifold planes are a common ingredient in 
phenomenologically interesting type II string theory,  
it seems natural to explore the possibility of treelevel 
de Sitter vacua or inflation models in type II string theory with 
orientifolds. On the other hand, there is no constraint for 
ekpyrotic scenario in type II string theory at present. 

The ekpyrosis inspired by string theory and brane world model 
suggests alternative solutions to the early universe puzzles 
such as inflation and dark energy, and assumes that two four-dimensional 
boundary branes which are located at the endpoints of 
orbifold in the higher-dimensional bulk spacetime 
\cite{Khoury:2001wf, Steinhardt:2001vw, 
Erickson:2006wc, Lehners:2008vx, Lehners:2009eg, Khoury:2009my}. 
For brane world picture, 
all forces except for gravity are localized on the branes while gravity 
can propagate freely in the bulk. When we assume that there 
is an attractive force between two branes, 
these branes approach to each other, which gives big bang. 
Since the big bang is described as a collision of branes 
there is not the beginning of time in the ekpyrotic scenario. 
Although two branes move through each other at once after collision, 
we can get a model so that the branes become closer again. 

The motion of branes is described by the potential of scalar field 
in a four-dimensional effective theory. 
Non-perturbative effects result in a potential which
attracts one brane towards the other brane \cite{Khoury:2001wf}.  
In order to resolve a horizon and flatness problem, the potential 
during a period of slow contraction before the big bang 
is negative and steeply falling. 
There was plenty of time before the big bang for the 
universe to be in causal contact over large 
regions, and in this way the horizon and flatness 
problem is automatically 
solved \cite{Lehners:2008vx}. Then, the usual statement of 
the ekpyrosis is that the universe slowly contracts before the 
big bang with the equation of state $w=p/\rho\gg 1$\,.   
The scalar potential is expected to turn up towards zero
at large negative value of scalar field in the ekpyrotic 
or cyclic models. 

As two branes approach each other, the branes are rippled because of 
generating quantum fluctuations. Since the collision of branes cannot 
be happened at exactly the same time, the branes collide slightly earlier or 
later in some places. Our universe thus has a little bit more time to cool or 
hot. When we consider the curvature perturbation in the ekpyrotic 
scenario, it occurs a strong blueshift, which is in sharp 
contradiction with the small redshift of the scalar fluctuation 
in the CMB \cite{Lyth:2001pf, Brandenberger:2001bs, 
Hwang:2001ga, Khoury:2001zk, Lyth:2001nv, Tsujikawa:2001ad, 
Notari:2002yc, Tsujikawa:2002qc, Lehners:2007ac}. 
The occurrence of blueshift scalar perturbation is nonetheless 
a minor flaw of ekpyrotic theory which can be easily corrected. 
Hence, new ekpyrotic theory considers multiple scalar field 
and successfully generated a scalar spectrum which is scale invariant and 
slightly redshifted 
\cite{Buchbinder:2007ad, Koyama:2007ag, Koyama:2007if, Buchbinder:2007at, 
Lehners:2007wc, Lehners:2008my, Mizuno:2008zza, Linde:2009mc, 
Lehners:2010fy, Li:2013hga, Battarra:2013cha, Fertig:2013kwa, Ijjas:2014fja,
Levy:2015awa, Fertig:2015ola}. 
Producing primordial gravitational waves sourced by the gauge field 
in the ekpyrotic scenario was also studied in \cite{Ito:2016fqp}. 

It is the purpose of this note to give a No-Go theorem of 
the ekpyrotic scenario in a ten-dimensional supergravity model 
which is low energy limit of a string theory.  
We study the dynamics of two scalar fields in the four-dimensional 
effective theory after a compactification in string theory. 
There are a dilaton and the volume modulus of the internal manifold 
in the effective theory, which is a four-dimensional theory of gravity 
minimally coupled to two-scalar fields.  
We will derive the two moduli fields with negative exponential potentials.
Since this potential is steep, the scalar field should be satisfied by 
``fast-roll'' condition instead of slow-roll parameter in inflation model 
\cite{Gratton:2003pe, Khoury:2003rt}. 
Although the terms coupling the scalar fields to the scalar curvature of the 
internal space and orientifold plane contribute the negative value to 
the potential in the string theory, we find that the potential does not 
satisfy the fast-roll condition in general. 
Therefore, it is not possible for us to realize the ekpyrotic phase 
in a string theory.  

Section \ref{sec:NOGO} describes the potential of scalar field for 
ekpyrotic scenario  
and the way it derives the four-dimensional 
effective theory. We discuss the approach to the effective action 
in more detail. The No-Go theorem of the ekpyrosis thus given by the
 string theory is discussed.  
We also investigate the detailed properties 
of these models, their embedding in a string theory and their 
viability. 
For simplicity, we do not consider D-branes and the associated moduli 
except for the volume modulus (breathing mode) 
of internal space although the analysis 
would not be different. 

Finally, section \ref{sec:Discussions} provides a brief summary
and an outlook to future developments. We have tried to make the context  
of this note relatively self-contained, but some details for the derivation 
of four-dimensional effective action is contained in appendix \ref{ap:sp}.

\section{No-Go theorem of the ekpyrotic scenario in the type II theory} 
\label{sec:NOGO}

In this section, we consider compactifications of the type II theory to 
four-dimensional spacetime on compact manifold Y. 
The ten-dimensional low-energy effective action for the type II theory takes 
the form \cite{Hertzberg:2007wc, Haque:2008jz}
\Eqr{
S&=&\frac{1}{2\bar{\kappa}^2}\int d^{10}x\sqrt{-g}
\left[\e^{-2\phi}\left(R+4g^{MN}\pd_M\phi\pd_N\phi
-\frac{1}{2}\left|H\right|^2\right)
-\frac{1}{2}\sum_p\left|F_p\right|^2\right]\nn\\
&&-\sum_p\left(T_{{\rm D}p}+T_{{\rm O}p}\right)
\int d^{p+1}x\sqrt{-g_{p+1}}\,\e^{-\phi}\,,
   \label{com:action:Eq}
}
where  
$\bar{\kappa}^2$ is the ten-dimensional gravitational constant, 
$R$ denotes the ten-dimensional Ricci scalar, $\phi$ is the scalar 
field, $H$ is the NS-NS 3-form field strength, $F_p$ are the R-R $p$-form 
field strengths ($p=0, 2, 4, 6, 8$ for type IIA, and 
$p=1, 3, 5, 7, 9$ for type IIB) that are sourced by D-branes, 
and $T_{{\rm D}p}$ $(T_{{\rm O}p})$ is the 
D$p$-brane (O$p$-plane) charge and tension.  
Although there are Chern-Simons terms in the ten-dimensional action, these are 
essentially independent of the dilaton and the scale of the background metric. 
Hence, we will not consider them. 

To compactify the theory to four dimensions, we consider the a  
metric ansatz of the form \cite{Hertzberg:2007wc, Haque:2008jz}
\Eqr{
ds^2&=&g_{MN}dx^Mdx^N=q_{\mu\nu}dx^\mu dx^\nu+
    g_{ij}dy^idy^j\nn\\
    &=&q_{\mu\nu}dx^\mu dx^\nu+
    \rho\,u_{ij}(\Ysp)dy^idy^j\,,
    \label{com:metric:Eq}
}
where $\rho$ is breathing mode (volume modulus of the compact space), 
$x^\mu$ denote the coordinates of four-dimensional spacetime, 
$y^i$ are local coordinates on the internal space Y, and 
$g_{MN}$\,, $q_{\mu\nu}$\,, $u_{ij}(\Ysp)$ are the metric of ten-dimensional 
spacetime, four-dimensional spacetime, six-dimensional internal space, 
respectively. 
We assume that $q_{\mu\nu}$\,, $u_{ij}$ depend only on the coordinates 
$x^\mu$, $y^i$\,, respectively. 
Since we factored out the overall volume modulus (breathing mode) 
$\rho$ of the internal space in the ten-dimensional metric 
(\ref{com:metric:Eq}), the modulus $\rho$ is related to the total 
physical volume of the internal space $v_6$ and the volume of Y space 
$v(\Ysp)$ as
\Eq{
\rho=\left[\frac{v_6}{v(\Ysp)}\right]^{1/3}\,,~~~~~
v_6=\int d^6y\sqrt{g_6}\,,~~~~~v(\Ysp)=\int d^6y\sqrt{u}\,.
}
Here, $g_6$\,, $u$ denote the determinant of the metric 
$g_{ij}$\,, $u_{ij}(\Ysp)$\,, respectively\,.  
The volume modulus $\rho$ is chosen such that the metric 
$u_{ij}(\Ysp)$ of the internal space is normalized $v(\Ysp)=1$\,. 

After we integrate over the internal space Y, the four-dimensional 
effective action $S_{\rm E}$ in the Einstein frame is given by 
\Eqr{
S_{\rm E}&=&\frac{1}{2\kappa^2}\int d^4x\sqrt{-\bar{q}}
\left[\bar{R}-\frac{3}{2}\bar{q}^{\mu\nu}\pd_\mu\ln \rho\,\pd_\nu\ln \rho
-2\bar{q}^{\mu\nu}\pd_\mu\ln \tau\,\pd_\nu\ln \tau\right.\nn\\
&&
+\left(\frac{\bar{\kappa}}{\tau\kappa}\right)^2\rho^{-1}R(\Ysp)
-\frac{1}{2}\left(\frac{\bar{\kappa}}{\tau\kappa}\right)^2\rho^{-3}
\left|H\right|^2
-\frac{1}{2}\sum_p\left(\frac{\bar{\kappa}}{\tau^2\kappa}\right)^{2}
\rho^{3-p}\left|F_p\right|^2\nn\\
&&\left.-2\sum_p\left(T_{{\rm D}p}+T_{{\rm O}p}\right)
\left(\frac{\bar{\kappa}^2}{\kappa}\right)^2\tau^{-3}\rho^{(p-6)/2}
\int d^{p-3}x\sqrt{g_{p-3}}\,\right],
   \label{co:ea:Eq}
}
where $R(\Ysp)$ denotes the Ricci scalar constructed from the metric 
$u_{ij}(\Ysp)$\, and $\kappa^2$ is the four-dimensional 
gravitational constant.  
Orientifold planes occupy $(p-3)$-dimensional internal space 
due to extending our four-dimensional universe. Then,   
the contribution of O$p$-plane~$(p\ge 3)$ to moduli potential 
will survive. 

In the four-dimensional action, we have defined the dilaton modulus 
\cite{Hertzberg:2007wc, Haque:2008jz}
\Eq{
\tau=\e^{-\phi}\rho^{3/2}\,,
}
and performed a conformal transformation on the four-dimensional metric
\Eq{
q_{\mu\nu}=\left(\frac{\bar{\kappa}}{\tau\kappa}\right)^2\bar{q}_{\mu\nu}\,.
    \label{co:4m:Eq}
}
Here, $\bar{q}_{\mu\nu}$ is the four-dimensional metric in the Einstein frame. 
$\bar{R}$ and $\bar{q}$ in the four-dimensional action (\ref{co:ea:Eq}) 
are the Ricci scalar and the determinant constructed 
from the metric $\bar{q}_{\mu\nu}$\,, respectively. 

Because of the conformal transformation (\ref{co:4m:Eq}), the kinetic term 
of the fields $\rho$ and $\tau$ is diagonal in the four-dimensional 
effective action. Since these fields do not have canonical kinetic energies, 
they are redefined as \cite{Hertzberg:2007wc}
\Eq{
\bar{\rho}=\sqrt{\frac{3}{2}}\kappa^{-1}\ln \rho\,,~~~~~
\bar{\tau}=\sqrt{2}\kappa^{-1}\ln \tau\,.
}
The moduli potential arises from the compactification of the terms 
in ten-dimensional action (\ref{com:action:Eq}) associated
with the various field strengths, D$p$-branes and O$p$-planes as well as 
the gravity and the dilaton. 
The 3-form $H$ and $p$-form field strengths 
$F_p$ can have a non-vanishing integral over
any closed three-, $p$-dimensional internal manifold of the 
compact space Y, and have to obey generalized Dirac charge 
quantization conditions, respectively
\Eq{
\int_{\Sigma}H=h_{\Sigma}\,,~~~~~
\int_{{\cal C}_p}F_p=f^{(p)}_{{\cal C}_p}\,,
    \label{co:Dirac:Eq} 
}
where $h_\Sigma$ and $f^{(p)}_{{\cal C}_p}$ are integers associated with 
number of quanta of $H$ and $F_p$ through each three-, $p$-dimensional 
homology cycles $\Sigma$\,, ${{\cal C}_p}$ in the internal manifold, 
respectively. 
We derive the potential energy in the four-dimensional Einstein frame 
arising from a three-form 
field strength $H$\,, and a $p$-form field strength $F_p$ 
coming from the terms in the ten-dimensional 
action (\ref{com:action:Eq}) which is proportional to 
$|H|^2$, $|F_p|^2$\,, respectively. 
We contract with three and $p$ factors of the internal space metric 
$g^{ij}$ so that 
\Eqrsubl{co:flux-p:Eq}{
&&V_H\propto \exp\left[-\kappa\left(\sqrt{2}\bar{\tau}
+\sqrt{6}\bar{\rho}\right)\right],~~{\rm for~~}H\,,\\
&&V_p\propto\exp\left[-\kappa\left\{2\sqrt{2}\bar{\tau}
+\frac{\sqrt{6}}{3}(p-3)\bar{\rho}\right\}\right],~~{\rm for~~}F_p\,.
}

Then we obtain the four-dimensional effective action in the Einstein frame: 
\Eqr{
S_{\rm E}
=\int d^4x\sqrt{-\bar{q}}
\left[\frac{1}{2\kappa^2}\bar{R}
-\frac{1}{2}\bar{q}^{\mu\nu}\pd_\mu\bar{\rho}\,\pd_\nu\bar{\rho}
-\frac{1}{2}\bar{q}^{\mu\nu}\pd_\mu\bar{\tau}\,\pd_\nu\bar{\tau}
-V(\bar{\tau}\,, \bar{\rho})\,\right],
   \label{co:ea2:Eq}
}
where the moduli potential of four-dimensional effective theory is 
given by 
\Eqr{
V(\bar{\tau}\,, \bar{\rho})=V_\Ysp+V_{\rm H}
+\sum_pV_p+\sum_pV_{{\rm D}p}+\sum_pV_{{\rm O}p}\,.
   \label{co:potential:Eq}
} 
Here, each components of the moduli potential can be expressed as 
\cite{Hertzberg:2007wc}
\Eqrsubl{co:p:Eq}{
V_\Ysp(\bar{\tau}\,, \bar{\rho})&=&-A_\Ysp
\exp\left[-\kappa\left(\sqrt{2}\bar{\tau}
+\frac{\sqrt{6}}{3}\bar{\rho}\right)\right]R(\Ysp)\,,\\
V_{\rm H}(\bar{\tau}\,, \bar{\rho})
&=&A_{\rm H}
\exp\left[-\kappa\left(\sqrt{2}\bar{\tau}
+\sqrt{6}\bar{\rho}\right)\right],\\
V_p(\bar{\tau}\,, \bar{\rho})&=&A_p
\exp\left[-\kappa\left\{2\sqrt{2}\bar{\tau}
+\frac{\sqrt{6}}{3}(p-3)\bar{\rho}\right\}\right],\\
V_{{\rm D}p}(\bar{\tau}\,, \bar{\rho})&=&A_{{\rm D}p}
\exp\left[-\kappa\left\{\frac{3\sqrt{2}}{2}\bar{\tau}
+\frac{\sqrt{6}}{6}(6-p)\bar{\rho}\right\}\right]
\int d^{p-3}x\sqrt{g_{p-3}}\,,\\
V_{{\rm O}p}(\bar{\tau}\,, \bar{\rho})&=&-
A_{{\rm O}p}
\exp\left[-\kappa\left\{\frac{3\sqrt{2}}{2}\bar{\tau}
+\frac{\sqrt{6}}{6}(6-p)\bar{\rho}\right\}\right] 
\int d^{p-3}x\sqrt{g_{p-3}}\,,
}
where $A_\Ysp$\,, $A_{\rm H}$\,, $A_p$\,, $A_{{\rm D}p}$\,, 
and $A_{{\rm O}p}$ are coefficients to scale with fluxes 
and numbers of O$p$-planes and D$p$-branes. These coefficients 
in general depend on the choice of flux integers $h_\Sigma$\,, 
$f^{(p)}_{{\cal C}_p}$\,, and 
also the function of the moduli of the internal space Y\,.

When the potential form for the ekpyrotic scenario gives the 
negative and steep, the fast-roll parameters for the 
ekpyrosis have to obey 
\cite{Gratton:2003pe, Khoury:2003rt}
\Eq{
\varepsilon_{\rm f}\equiv\kappa^2\frac{V^2}{\left(\pd_{\bar{\tau}}V\right)^2
+\left(\pd_{\bar{\rho}}V\right)^2}\ll 1\,,~~~~~
\left|\eta_{\rm f}\right|\equiv 
\left|1-\frac{V\left(\pd^2_{\bar{\tau}}V
+\pd^2_{\bar{\rho}}V\right)}{\left(\pd_{\bar{\tau}}V\right)^2
+\left(\pd_{\bar{\rho}}V\right)^2}\right|\ll 1\,. 
  \label{co:frp:Eq}
}
This is analogy with the standard slow-roll parameters in inflation. 
The potential form satisfying the condition (\ref{co:frp:Eq}) 
gives the ekpyrotic period of slow contraction before the big bang. 

In the following, we illustrate how the above ingredients may be 
useful from the point of view of excluding the ekpyrosis. 
We focus the discussion on the contribution of negative energy,
which appear particularly promising. 
Our goal is only to show that simple, available 
ingredients in the type II theory have energy densities which scale 
with the volume and dilaton moduli in a way which suffices to 
obey our no-go theorem, which was based purely on scaling of energy 
densities. This should act as a guide to model building,
but should be taken in the heuristic spirit it is offered.

It is especially interesting to understand the dynamics of 
moduli at negative potential energy.  
If there are non-trivial fluxes in the background (\ref{co:p:Eq}), 
one notes that these make uplifting the moduli potential to positive energy.  
In this note, we consider the moduli potential (\ref{co:ea2:Eq})  
without flux:
\Eqr{
\hspace{-1cm}
V(\bar{\tau}\,, \bar{\rho})&=&V_\Ysp+\sum_pV_{{\rm O}p}\nn\\
\hspace{-1cm}&=&-A_\Ysp
\exp\left[-\kappa\left(\sqrt{2}\bar{\tau}
+\frac{\sqrt{6}}{3}\bar{\rho}\right)\right]R(\Ysp)\nn\\
\hspace{-1cm}&&-\sum_p
A_{{\rm O}p}
\exp\left[-\kappa\left\{\frac{3\sqrt{2}}{2}\bar{\tau}
+\frac{\sqrt{6}}{6}(6-p)\bar{\rho}\right\}\right]
\int d^{p-3}x\sqrt{g_{p-3}}\,.
}

Before discussing the No-Go theorem, we comment about the other moduli 
of the theory associated with the compactification. The four-dimensional 
effective action (\ref{co:ea2:Eq}) includes in general 
kinetic energy terms of so-called 
K\"ahler moduli, complex structure moduli, and axions. 
As we have mentioned above, the various 
coefficients $A_\Ysp$\,, $A_{\rm H}$\,, $A_p$\,, $A_{{\rm D}p}$\,, 
and $A_{{\rm O}p}$ in the potential (\ref{co:p:Eq}) 
are complicated functions of these moduli. 
Although there are kinetic terms of these moduli in the four-dimensional 
effective action, their contributions will always be positive 
\cite{Hertzberg:2007wc}. For simplicity, we do not consider the dynamics and 
fixing of these moduli in the following. 
However, if four-dimensional effective action is described by 
K\"ahler moduli, complex structure moduli, and axions as well as 
volume moduli (breathing mode), 
the moduli potential will be modified. 
We will discuss these in the end of this 
section. The stabilization mechanisms of all the
geometric moduli and many axions in type II string theory 
have been discussed in \cite{Giddings:2001yu, Kachru:2003aw, 
Villadoro:2005cu, DeWolfe:2005uu}. 

\subsection{Type IIA compactification}
\label{sec:IIA}

There are No-Go theorems which exclude 
slow roll inflation and de Sitter vacua in the 
simple IIA compactifications with orientifold planes 
\cite{Hertzberg:2007wc, deCarlos:2009fq}. 
For the IIA flux compactifications on Calabi-Yau manifolds with O6-planes 
in the four-dimensional potential (\ref{co:potential:Eq}), the 
slow-roll inflation is forbidden \cite{Hertzberg:2007wc}. 
The construction classical de Sitter vacua 
in IIA compactifications on internal manifolds with negative curvature
and orientifold planes were studied in 
\cite{Haque:2008jz, Silverstein:2007ac, Danielsson:2009ff}. 

\begin{figure}[h]
 \begin{center}
\includegraphics[keepaspectratio, scale=0.6]{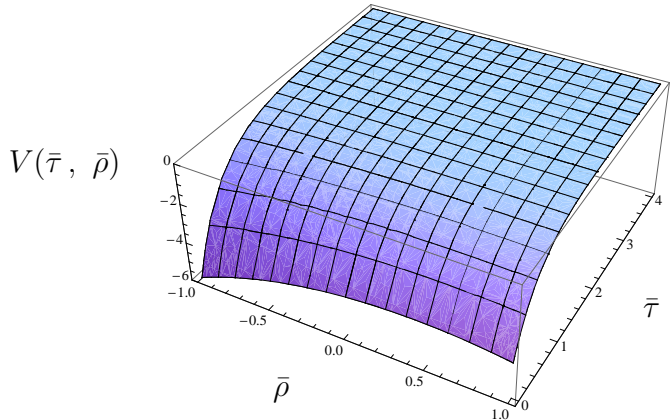}
\put(-250,95){$V(\bar{\tau}\,,~\bar{\rho})$}
\put(-10,40){$\bar{\tau}$}
\put(-150,10){$\bar{\rho}$}\\
  \caption{\baselineskip 14pt
We depict the 
moduli potential $V(\bar{\tau}\,,~\bar{\rho})$ in the type IIA 
theory for the case of $R(\Ysp)=1$\,, 
$A_\Ysp=1$\,, $A_{\rm H}=A_p=A_{{\rm D}p}
=0$\,, and $A_{{\rm O}p}\int d^{p-3}x\sqrt{g_{p-3}}=1\,,~
(p=4, 6, 8)$\,. We also fix $\kappa=1$\, \cite{Hertzberg:2007wc}. 
The moduli potential has negative value.  
}
  \label{fig:potential}
 \end{center}
\end{figure}

\begin{figure}[h]
 \begin{center} 
\includegraphics[keepaspectratio, scale=0.6]{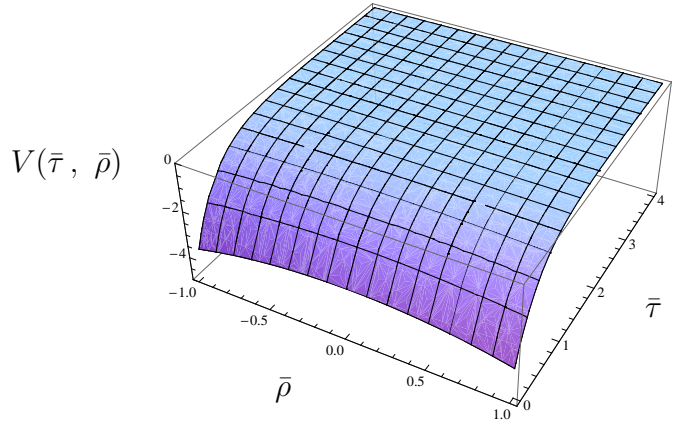}
\put(-250,95){$V(\bar{\tau}\,,~\bar{\rho})$}
\put(-10,40){$\bar{\tau}$}
\put(-150,10){$\bar{\rho}$}\\
  \caption{\baselineskip 14pt 
The moduli potential $V(\bar{\tau}\,,~\bar{\rho})$ 
in the type IIA theory for the case of 
$R(\Ysp)=0$\,, $A_{\rm H}=A_p=A_{{\rm D}p}
=0$\,, and $A_{{\rm O}p}\int d^{p-3}x\sqrt{g_{p-3}}=1\,,~
(p=4, 6, 8)$ is depicted.  
Fixing the four-dimensional gravitational constant as $\kappa=1$\,, 
we obtain the negative potential. 
}
  \label{fig:potential2}
 \end{center}
\end{figure}

In this section, we consider IIA compactifications on an internal space 
(\ref{co:4m:Eq}), namely positive curvature and Ricci flat spaces, 
involving orientifold planes, and discuss the No-Go theorem for 
ekpyrotic scenario. 
The analysis will focus on the behavior of the moduli potential 
in the volume modulus and dilaton. 
In order to present the no-go theorem using these fields, 
we have to still make sure that there are 
no steep directions of the scalar potential in the ($\bar{\rho}$\,, 
$\bar{\tau}$)-plane. The scalar potentials are shown in 
Fig.~\ref{fig:potential} for $R(\Ysp)=1$, and Fig.~\ref{fig:potential2} 
for $R(\Ysp)=0$\,. 

\begin{figure}[h]
 \begin{center}
\includegraphics[keepaspectratio, scale=0.5]{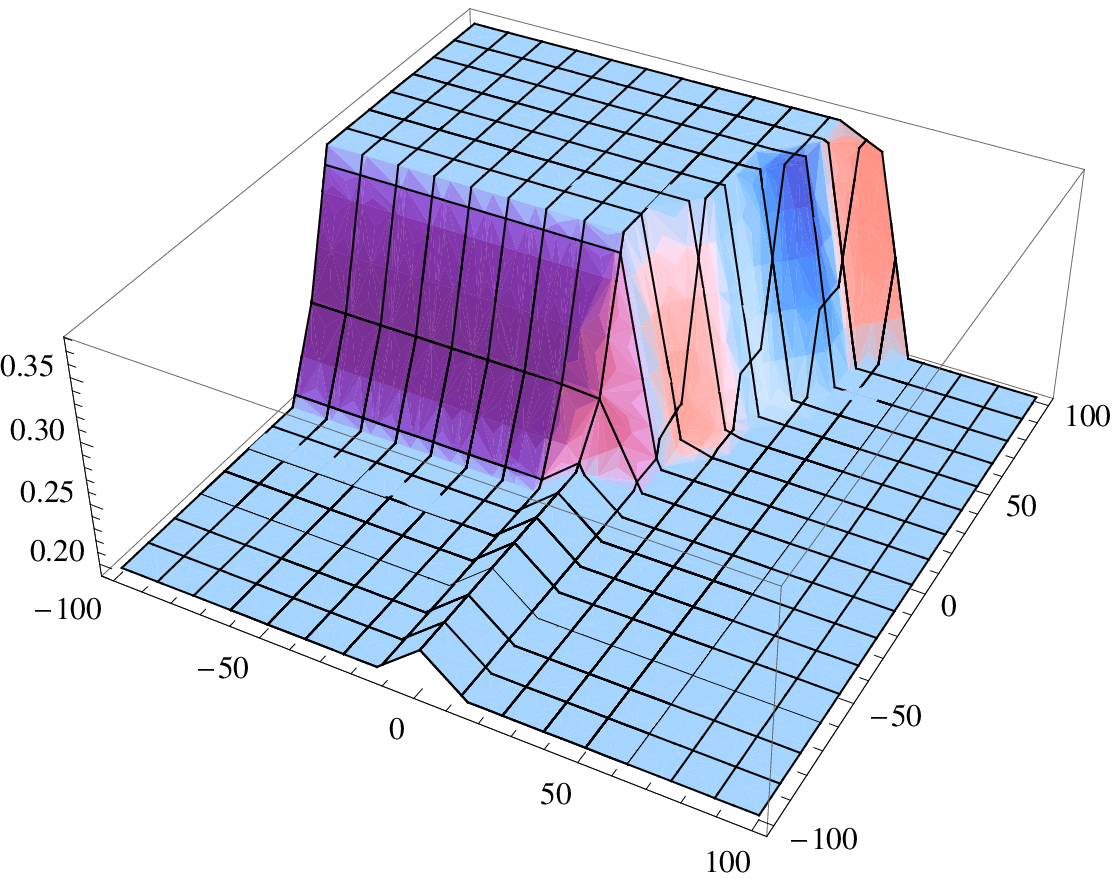}
\put(-185,90){$\varepsilon_{\rm f}(\bar{\tau}\,,~\bar{\rho})$}
\put(0,30){$\bar{\tau}$}
\put(-120,5){$\bar{\rho}$}
\hskip 2cm
\includegraphics[keepaspectratio, scale=0.5]{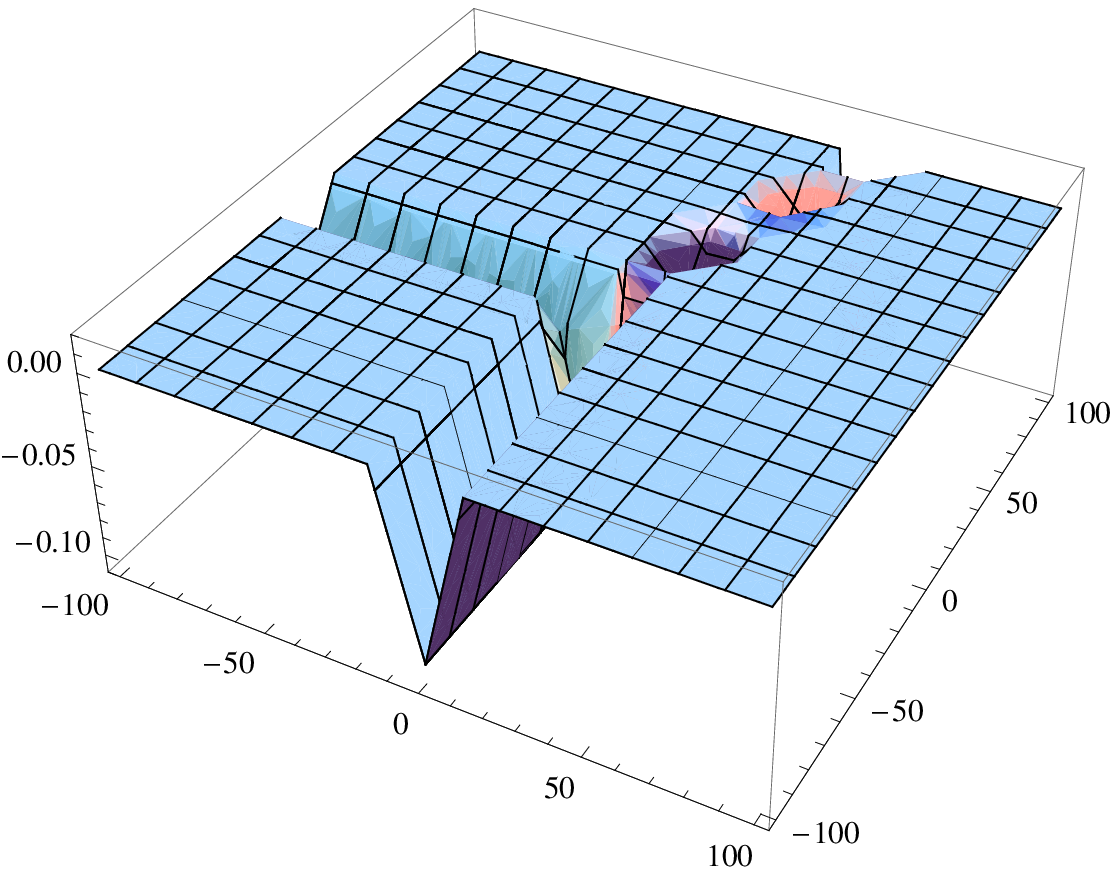}
\put(-185,90){$\eta_{\rm f}(\bar{\tau}\,,~\bar{\rho})$}
\put(0,30){$\bar{\tau}$}
\put(-120,5){$\bar{\rho}$}\\
(a) \hskip 7cm (b)  ~~~~~~
  \caption{\baselineskip 14pt 
For the case of $R(\Ysp)=1$\,,
$A_\Ysp=1$\,, $A_{\rm H}=A_p=A_{{\rm D}p}
=0$\,, and $A_{{\rm O}p}\int d^{p-3}x\sqrt{g_{p-3}}=1\,,~
(p=4, 6, 8)$\,, in the moduli potential (\ref{co:p:Eq}) 
of the type IIA theory, the parameters 
$\varepsilon_{\rm f}(\bar{\tau}\,,~\bar{\rho})$\,, 
$\eta_{\rm f}(\bar{\tau}\,,~\bar{\rho})$\,, are depicted. 
We set $\kappa=1$\,\cite{Hertzberg:2007wc}. 
Since these results give the contradiction 
with the fast-roll condition for ekpyrosis (\ref{co:frp:Eq})\,,  
the ekpyrosis is not allowed. 
}
  \label{fig:ep-et}
 \end{center}
\end{figure}

\begin{figure}[h]
 \begin{center}
\includegraphics[keepaspectratio, scale=0.5]{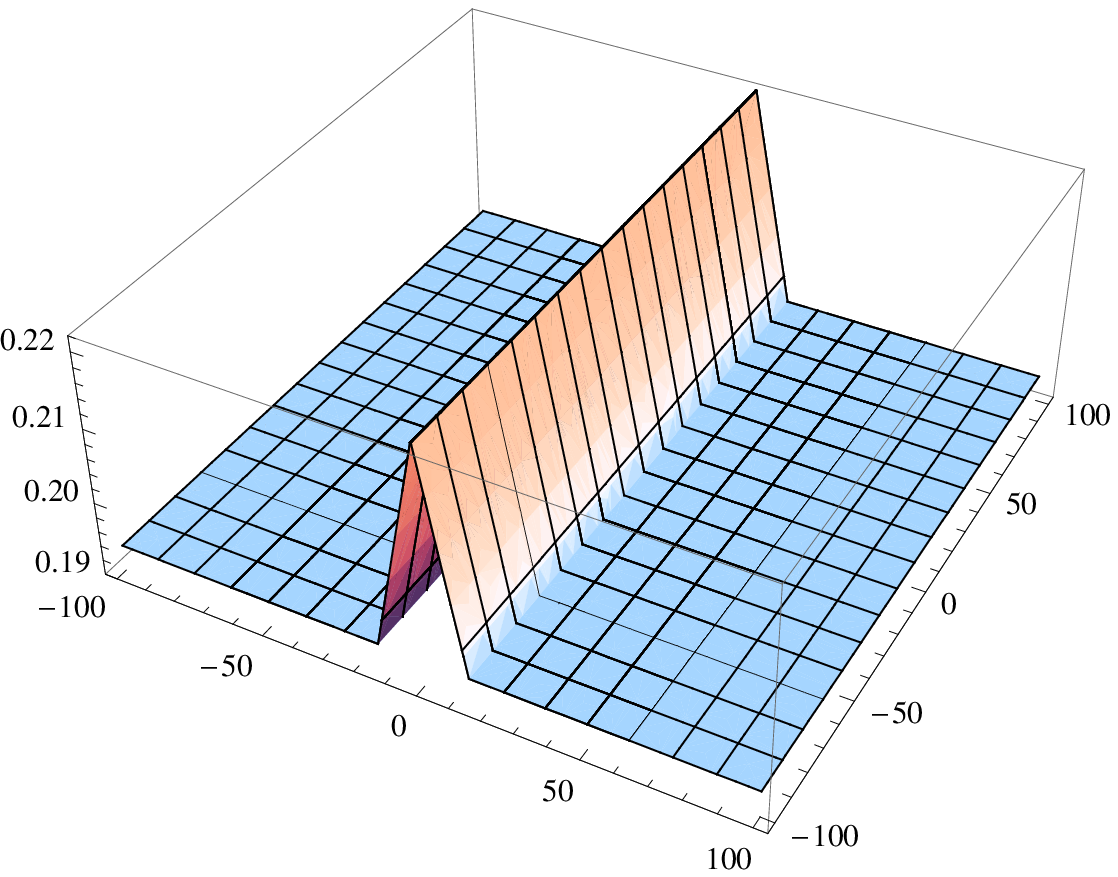}
\put(-185,90){$\varepsilon_{\rm f}(\bar{\tau}\,,~\bar{\rho})$}
\put(0,30){$\bar{\tau}$}
\put(-120,5){$\bar{\rho}$}
\hskip 2cm
\includegraphics[keepaspectratio, scale=0.5]{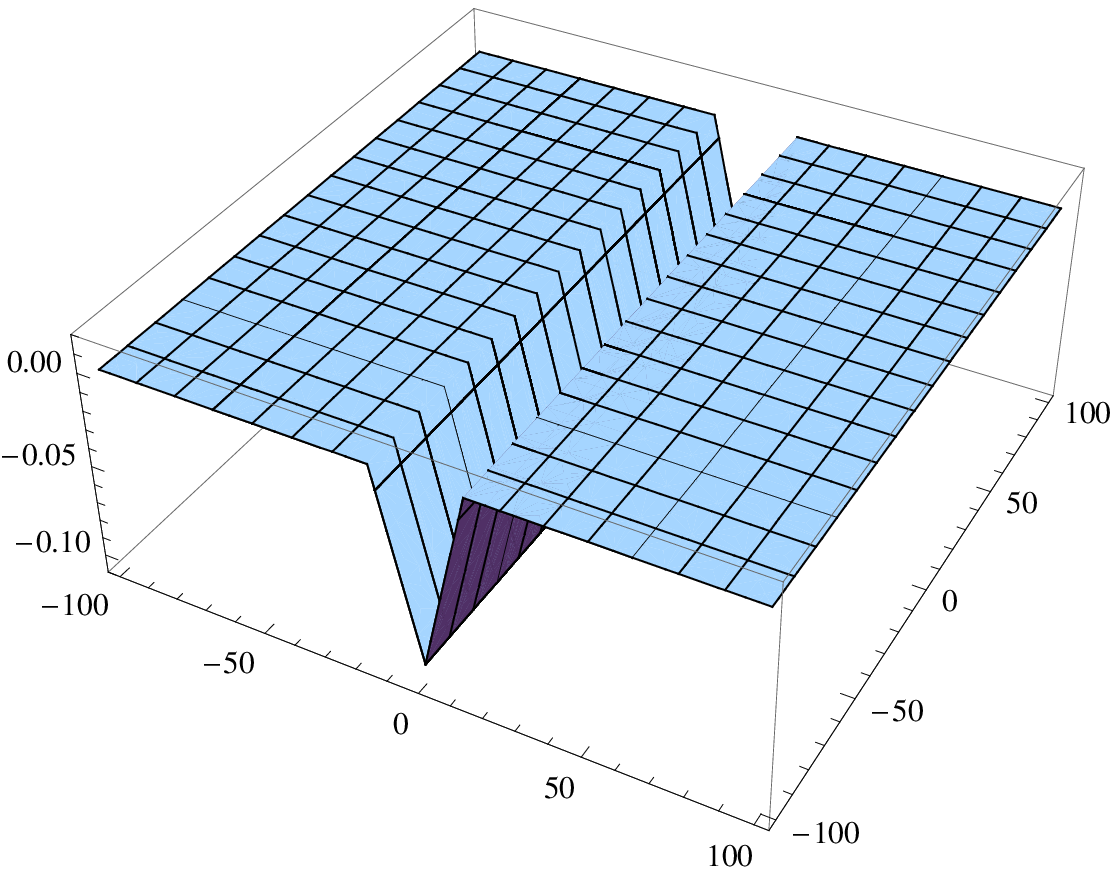}
\put(-185,90){$\eta_{\rm f}(\bar{\tau}\,,~\bar{\rho})$}
\put(0,30){$\bar{\tau}$}
\put(-120,5){$\bar{\rho}$}\\
(a) \hskip 7cm (b)  ~~~~~~
  \caption{\baselineskip 14pt 
We fix 
$R(\Ysp)=0$\,,  $A_{\rm H}=A_p=A_{{\rm D}p}
=0$\,, and $A_{{\rm O}p}\int d^{p-3}x\sqrt{g_{p-3}}=1\,,~
(p=4, 6, 8)$\,, and set $\kappa=1$\,. 
in the moduli potential (\ref{co:p:Eq}) for the type IIA theory. 
We depicted the behavior of 
the parameters 
$\varepsilon_{\rm f}(\bar{\tau}\,,~\bar{\rho})$\,, 
$\eta_{\rm f}(\bar{\tau}\,,~\bar{\rho})$\,.  
Our setup cannot describe the ekpyrotic scenario 
because the background does not obey the fast-roll condition 
(\ref{co:frp:Eq}).}
  \label{fig:ep-et2}
 \end{center}
\end{figure}

In such cases one can then study directions involving
$\bar{\rho}$\,,  
$\bar{\tau}$ and finds that the scalar potential satisfies 
\Eqr{
\varepsilon_{\rm f}=\frac{V^2}{2}\left[
\left\{V_\Ysp+\frac{3}{2}\left(
V_{\rm O4}+V_{\rm O6}+V_{\rm O8}\right)\right\}^2
+\frac{1}{3}\left(V_\Ysp+V_{\rm O4}-V_{\rm O8}\right)^2
\right]^{-1}>\frac{6}{31}\,.
}
The fast-roll parameter $\varepsilon_{\rm f}$ has the bound 
$\varepsilon_{\rm f}> 6/31$\,. This results does not depend on  
the choice of coefficients $A_\Ysp$ and $A_{{\rm O}p}$\,.  
The value of parameter $\varepsilon_{\rm f}$ is not much less than one, 
which is the contradiction 
with the fast-roll condition for ekpyrosis (\ref{co:frp:Eq})\,. 
Hence, ekpyrosis is not allowed. 
This can also be seen from the results in 
\cite{Meeus:2016}.  
We illustrate the configuration of parameters 
$\varepsilon_{\rm f}$\,, $\eta_{\rm f}$ in Figs.~\ref{fig:ep-et}, 
(for $R(\Ysp)>0$) and \ref{fig:ep-et2} (for $R(\Ysp)=0$)\,. 

\subsection{Type IIB compactification}
\label{sec:IIB}

There are many no-go theorems and exclude most concrete
examples of de Sitter solution \cite{Giddings:2001yu, Gibbons:1984kp, 
deWit:1986mwo, Maldacena:2000mw, Caviezel:2009tu}. 
It is possible to evade simple no-go 
theorems of inflationary scenario for SU(2)$\times$SU(2) 
with an SU(2)-structure and O5- and O7-planes. Although  
there are de Sitter critical points, these have at least one tachyonic 
direction with a large $\eta$ (slow-roll) parameter 
\cite{Caviezel:2009tu}. 

\begin{figure}[h]
 \begin{center} 
\includegraphics[keepaspectratio, scale=0.6]{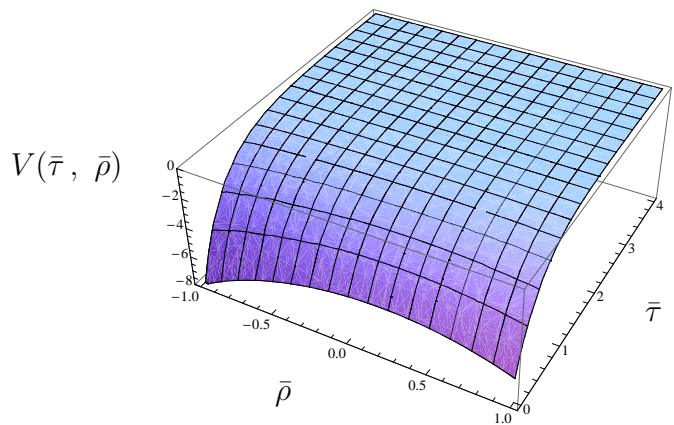}
\put(-250,95){$V(\bar{\tau}\,,~\bar{\rho})$}
\put(-10,40){$\bar{\tau}$}
\put(-150,10){$\bar{\rho}$}\\
  \caption{\baselineskip 14pt 
The figure shows the 
moduli potential $V(\bar{\tau}\,,~\bar{\rho})$ 
in the type IIB theory for the case of $R(\Ysp)=1$\,, 
$A_\Ysp=1$\,, $A_{\rm H}=A_p=A_{{\rm D}p}
=0$\,, $A_{{\rm O}p}\int d^{p-3}x\sqrt{g_{p-3}}=1\,,~
(p=3, 5, 7, 9)$\,, and $\kappa=1$\,.  
The potential of the moduli becomes negative without flux.  
}
  \label{fig:potentialb}
 \end{center}
\end{figure}

\begin{figure}[h]
 \begin{center}
\includegraphics[keepaspectratio, scale=0.6]{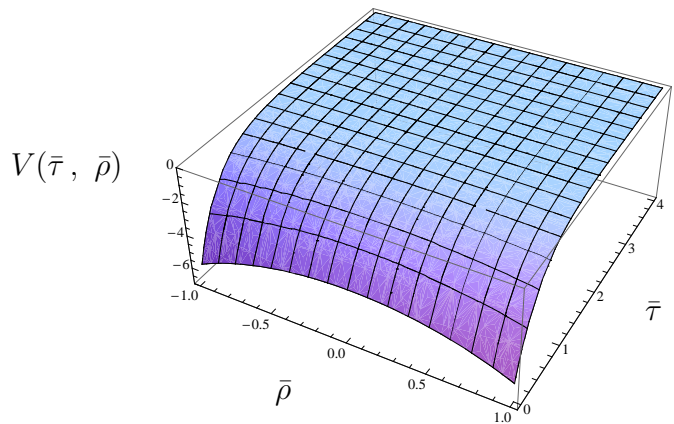}
\put(-250,95){$V(\bar{\tau}\,,~\bar{\rho})$}
\put(-10,40){$\bar{\tau}$}
\put(-150,10){$\bar{\rho}$}\\
  \caption{\baselineskip 14pt 
The moduli potential $V(\bar{\tau}\,,~\bar{\rho})$ 
in the type IIB theory for the case of 
$R(\Ysp)=0$\,, $A_{\rm H}=A_p=A_{{\rm D}p}
=0$\,, and $A_{{\rm O}p}\int d^{p-3}x\sqrt{g_{p-3}}=1\,,~
(p=3, 5, 7, 9)$\,, and $\kappa=1$ 
is shown in the $(\bar{\tau}\,,~\bar{\rho})$ space. 
Although the moduli potential becomes negative, the form 
of the moduli potential is not steep due to $\varepsilon_{\rm f}>1/6$\,,~
 (See Fig.~\ref{fig:ep-etb2})\,. In this setup, we cannot 
describe the ekpyrotic scenario. 
}
  \label{fig:potentialb2}
 \end{center}
\end{figure}

On the other hand, 
for type IIB compactifications in the ekpyrotic model, 
we have also seen that it is possible to obtain simple no-go 
theorems in the ($\bar{\rho}$\,, $\bar{\tau}$)-plane if 
one includes orientifold planes 
and the curvature of the internal space.  The behavior of four-dimensional 
scalar potential is numerically illustrated in Fig.~\ref{fig:potentialb} for 
$R(\Ysp)=1$, and in Fig.~\ref{fig:potentialb2} for $R(\Ysp)=0$\,. 

From the Eq.~(\ref{co:frp:Eq}), we find a constraint of 
the fast-roll parameter $\varepsilon_{\rm f}$  
\Eqr{
\varepsilon_{\rm f}&=&V^2\left[
2\left\{V_\Ysp+\frac{3}{2}\left(
V_{\rm O3}+V_{\rm O5}+V_{\rm O7}+V_{\rm O9}\right)\right\}^2
\right.\nn\\
&&\left.+\frac{1}{6}\left(2V_\Ysp+3V_{\rm O3}+V_{\rm O5}
-V_{\rm O7}-3V_{\rm O9}\right)^2
\right]^{-1}>\frac{1}{6}\,.
}

\begin{figure}[h]
 \begin{center}
\includegraphics[keepaspectratio, scale=0.5]{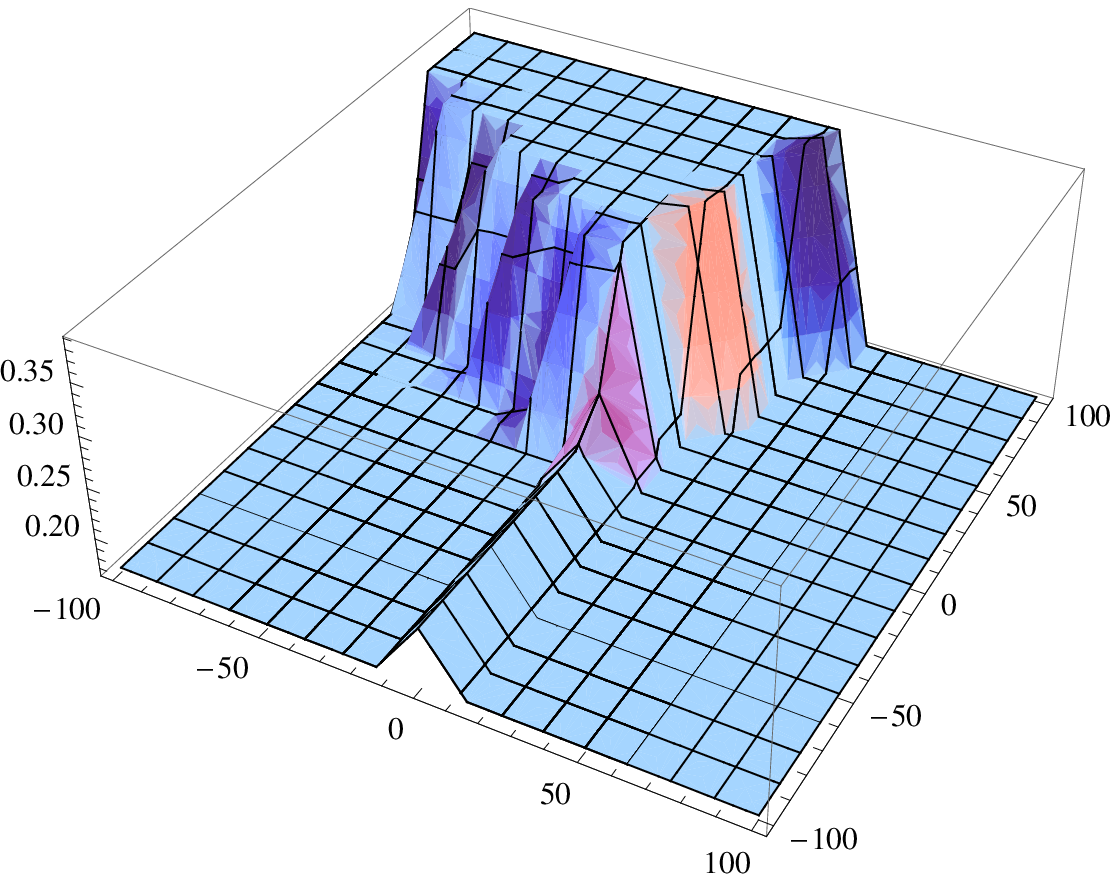}
\put(-185,90){$\varepsilon_{\rm f}(\bar{\tau}\,,~\bar{\rho})$}
\put(0,30){$\bar{\tau}$}
\put(-120,5){$\bar{\rho}$}
\hskip 2cm
\includegraphics[keepaspectratio, scale=0.5]{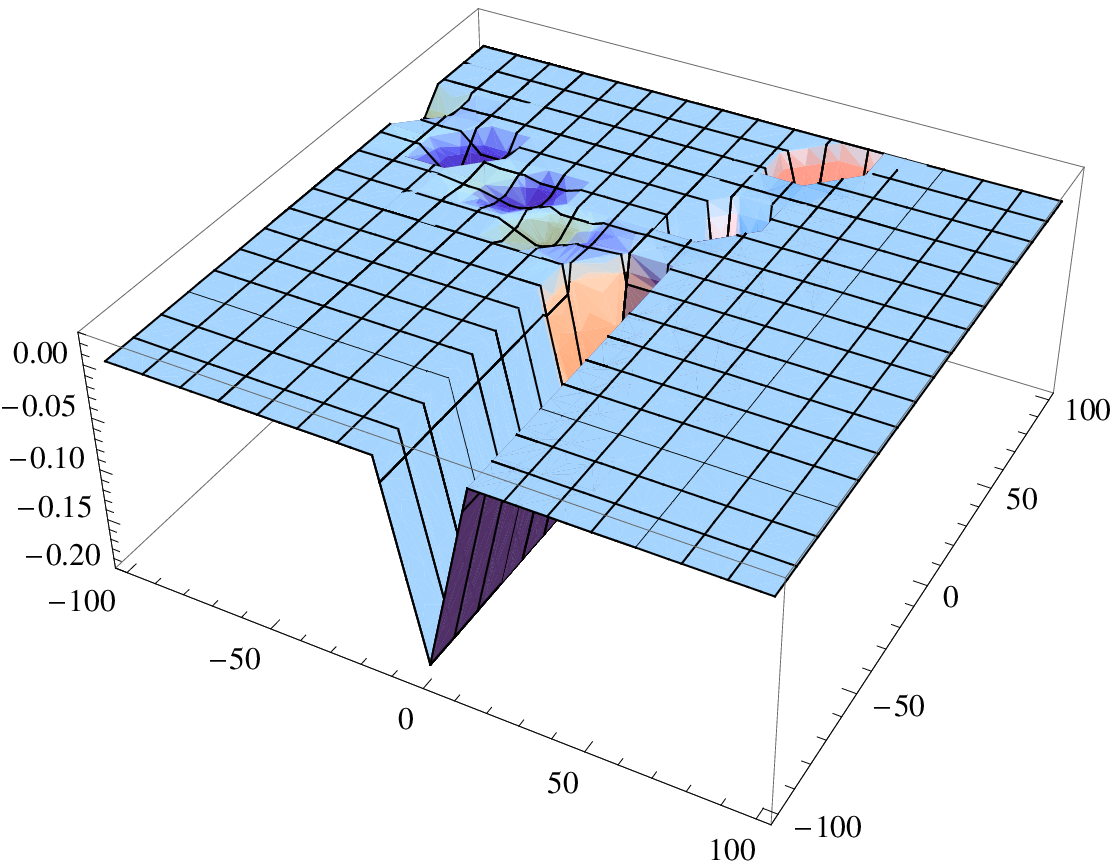}
\put(-185,90){$\eta_{\rm f}(\bar{\tau}\,,~\bar{\rho})$}
\put(0,30){$\bar{\tau}$}
\put(-120,5){$\bar{\rho}$}\\
(a) \hskip 7cm (b)  ~~~~~~
  \caption{\baselineskip 14pt
We shows the behavior of parameters
$\varepsilon_{\rm f}(\bar{\tau}\,,~\bar{\rho})$\,, 
$\eta_{\rm f}(\bar{\tau}\,,~\bar{\rho})$ for the type IIB theory. 
We set $R(\Ysp)=1$\,, $A_\Ysp=1$\,, $A_{\rm H}=A_p=A_{{\rm D}p}
=0$\,, $A_{{\rm O}p}\int d^{p-3}x\sqrt{g_{p-3}}=1\,,~
(p=3, 5, 7, 9)$\,, and $\kappa=1$\,, in the moduli potential 
(\ref{co:p:Eq}), the parameters 
$\varepsilon_{\rm f}(\bar{\tau}\,,~\bar{\rho})$\,, 
$\eta_{\rm f}(\bar{\tau}\,,~\bar{\rho})$\,, are depicted. 
Since these results are not consistent with  
the fast-roll condition for ekpyrosis (\ref{co:frp:Eq})\,,   
we cannot describe the ekpyrotic scenario in our background.
}
  \label{fig:ep-etb}
 \end{center}
\end{figure}

\begin{figure}[h]
 \begin{center}
\includegraphics[keepaspectratio, scale=0.5]{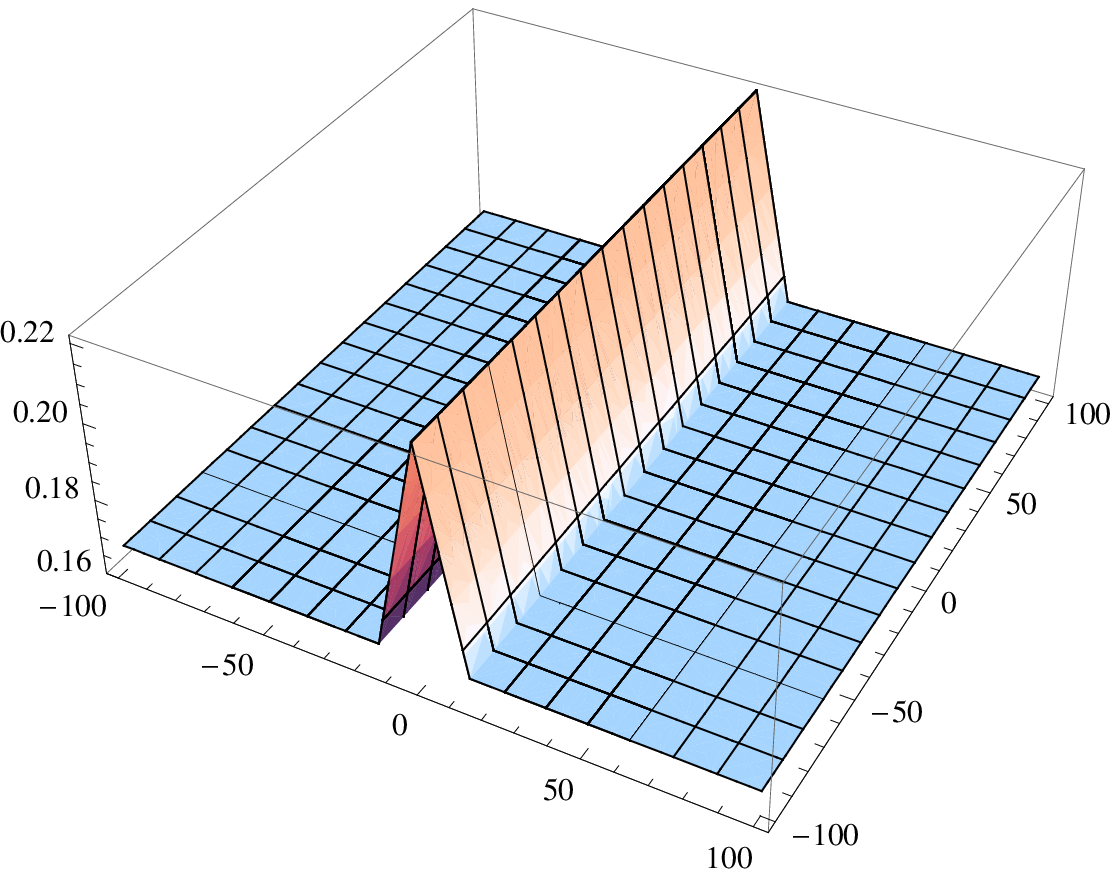}
\put(-185,90){$\varepsilon_{\rm f}(\bar{\tau}\,,~\bar{\rho})$}
\put(0,30){$\bar{\tau}$}
\put(-120,5){$\bar{\rho}$}
\hskip 2cm
\includegraphics[keepaspectratio, scale=0.5]{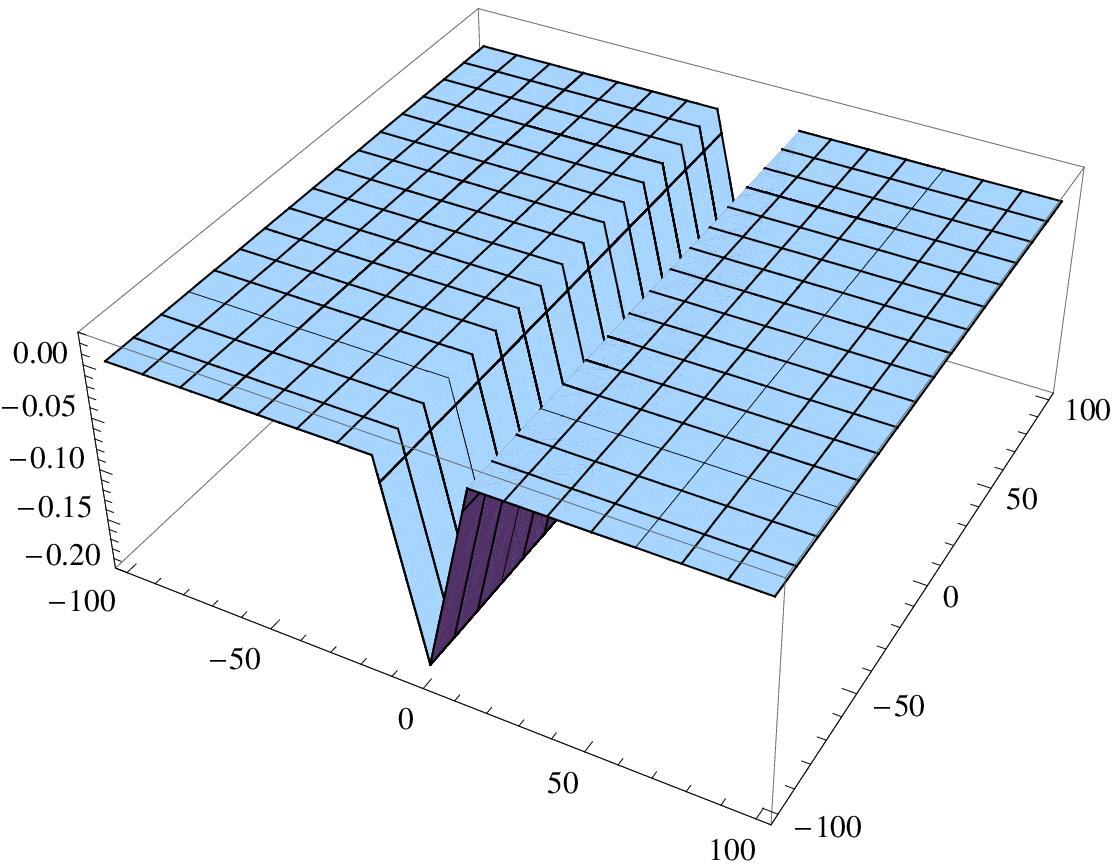}
\put(-185,90){$\eta_{\rm f}(\bar{\tau}\,,~\bar{\rho})$}
\put(0,30){$\bar{\tau}$}
\put(-120,5){$\bar{\rho}$}\\
(a) \hskip 7cm (b)  ~~~~~~
  \caption{\baselineskip 14pt 
We show the parameters $\varepsilon_{\rm f}(\bar{\tau}\,,~\bar{\rho})$\,, 
and $\eta_{\rm f}(\bar{\tau}\,,~\bar{\rho})$\,, numerically 
for the type IIB theory. 
Our parameters are $R(\Ysp)=0$\,,  $A_{\rm H}=A_p=A_{{\rm D}p}
=0$\,, and $A_{{\rm O}p}\int d^{p-3}x\sqrt{g_{p-3}}=1\,,~
(p=3, 5, 7, 9)$\,, and $\kappa=1$\,, 
in the moduli potential (\ref{co:p:Eq}). It shows that our setup 
does not satisfy the fast-roll condition (\ref{co:frp:Eq}).  
}
  \label{fig:ep-etb2}
 \end{center}
\end{figure}

Unfortunately, the form moduli potential is not steep again as 
$\varepsilon_{\rm f}$ and $\eta_{\rm f}$ parameters turns 
out to be large value. Just as in the IIA analogue, 
one obtains the bound $\varepsilon_{\rm f}> 1/6$\,. 
If we choose different values for $A_\Ysp$ and $A_{{\rm O}p}$  
in the moduli potential (\ref{co:p:Eq}), we can find  
again the same bound. 
We show fast-roll parameters $\varepsilon_{\rm f}$ and 
$\eta_{\rm f}$ numerically in Fig.~\ref{fig:ep-etb} 
(for $R(\Ysp)=1$) and 
Fig.~\ref{fig:ep-etb2} (for $R(\Ysp)=0$) in the IIB theory.

If we consider the dynamics of remaining moduli 
such as K\"ahler moduli, complex structure moduli and axions,  
the potential is in general a function of hundreds of fields, 
for large number of infinite families of possible flux combinations 
on each of the many available internal manifolds. 
When we treat dynamics of these moduli or    
the coupling between moduli and fluxes or D-brane, O-plane,  
they contribute the potential. 
There is a possibility that the no-go theorems using these fields 
are circumvented.  
%
\section{Discussions}
  \label{sec:Discussions}
In this note, we have studied the No-Go theorem of the ekpyrosis for 
string theory in a spacetime of ten dimensions. We gave a potential  
of the scalar fields in four-dimensional 
effective theory, in terms of the compactification with smooth manifold. 

The effective potential of two scalar fields can be constructed 
by postulating suitable emergent gravity, orientifold planes, 
and vanishing fluxes on the ten-dimensional background. 
The construction of four-dimensional effective action was developed in 
\cite{Hertzberg:2007wc}, mainly but not entirely in the context of a 
new ekpyrotic scenario. 

We have used the results of section \ref{sec:NOGO} to analyze 
the dynamics of moduli that can be constructed using a simple 
compactification. The scalar potential depends only on two 
moduli: $\bar{\rho}$ and $\bar{\tau}$\,. 
In such a simple setting, one can show that $\varepsilon_{\rm f}>6/31$ 
for IIA theory and $\varepsilon_{\rm f}>1/6$ for IIB theory 
whenever $V(\bar{\rho}\,, \bar{\tau})<0$\,. 
It has been known for some time \cite{Lehners:2008vx} that the effective 
potential of scalar fields requires the fast-roll parameter to be small 
during the ekpyrotic phase. However, with the help of the tools developed 
in section \ref{sec:NOGO}, this is prohibited in a string theory with 
a compactification. 
Hence, the explicit nature of the dynamics has made it impossible 
to realize the ekpyrotic phase in the present note. This is 
consistent with the results in \cite{Meeus:2016}. 

As we have commented in Sec.~\ref{sec:NOGO}\,, we have not considered 
the dynamics for moduli other than the  
volume modulus of the six-dimensional internal space 
in this note. If these moduli have dynamics, 
kinetic terms of moduli appear in the four-dimensional 
effective theory. Moreover, since moduli couple to orientifolds,  
it is possible that there are other steep directions of the scalar 
potential in directions outside the ($\bar{\rho}$\,, $\bar{\tau}$)-plane. 

The compactification we have considered in this note also gives the 
No-Go theorem of the inflationary scenario as well as the ekpyrosis because
the moduli potential cannot satisfy the slow-roll condition 
\cite{Hertzberg:2007wc}. There are constrains or No-Go theorems to 
construct the inflationary or de Sitter model in the string theory 
with a few simple assumptions 
\cite{Hertzberg:2007wc, Danielsson:2009ff, Wrase:2010ew, 
Shiu:2011zt, VanRiet:2011yc}.  

In order to embed ekpyrotic or cyclic models in a ten-dimensional 
supergravity we have investigated in this note, we may  
consider some ingredients, for instance, the 
dynamics of remaining moduli, higher curvature correction 
other than orientifold and flux. 
We have not attempted an explicit construction here, 
since that will take us beyond the scope of this note. 
A lot of study remains to be done 
in string theory before a cosmologically realistic case is treated. 

\section*{Acknowledgments}
We would like to thank Thomas Van Riet for numerous valuable 
discussions and careful reading of the manuscript, and  
thank Tetsuya Shiromizu, Masato Minamitsuji, Shuntaro Mizuno, 
Toshifumi Noumi, Yusuke Yamada, Taishi Ikeda, Asuka Ito for discussions 
and valuable comments. We would like to also 
thank the Yukawa Institute for Theoretical Physics at Kyoto University.
Discussions during the YITP symposium YKIS2018a ``General Relativity -- 
The Next Generation --'' and the workshop YITP-T-17-02 ``Gravity and 
Cosmology 2018'' were useful to complete this work. We are 
grateful to Shinji Mukohyama and Atsushi Watanabe for their warm 
hospitalities in the Yukawa Institute for Theoretical Physics. 
We also thank the Yukawa Institute for Theoretical Physics at Kyoto 
University for hospitality during the YITP workshop on gGravity and Cosmology 
for Young Researchersh (Workshop No. YITP-X-17-08) which was supported 
by Grant-in-Aid for Scientific Research on Innovative Areas No. 17H06359. 
This work is supported by Grants-in-Aid from the Scientific Research 
Fund of the Japan Society for the Promotion of Science, 
under Contracts No. 16K05364.

\appendix

\section{Scalar potential in four-dimensional effective theory}
\label{ap:sp}

In this appendix, we show how to find the scalar potential in four-dimensional 
effective theory.  
 We present the explicit procedure for the cases of type II theory 
with D$p$-brane and O$p$-plane systems. 

We assume that ten-dimensional 
action and metric are given by (\ref{com:action:Eq}), 
(\ref{com:metric:Eq}), respectively. 
We first check the gravity sector. 
Upon setting the ten-dimensional 
metric (\ref{com:metric:Eq}), we find 
\Eq{
\frac{1}{2\bar{\kappa}^2}\int d^{10}x\sqrt{-g}\,\e^{-2\phi}\,R
=\frac{1}{2\bar{\kappa}^2}\int d^4x\sqrt{-q}\,\e^{-2\phi}\,R\,\rho^3
\int d^6y \sqrt{u}\,,
}
where $u$ denotes the determinant of the metric $u_{ij}(\Ysp)$ 
in (\ref{com:metric:Eq})\,.  
We have used the volume modulus of the compact internal space $\rho$ and 
defined the dilaton modulus $\tau$ by
\Eq{
\rho=\left[\frac{\int d^6y\sqrt{g_6}}{\int d^6y\sqrt{u}}\right]^{1/3}\,, 
~~~~~\tau=\e^{-\phi}\rho^{3/2}\,,
}
where $g_6$ is the determinant of the metric $g_{ij}$\,.  
In order to obtain the canonical form in the gravity sector, 
we perform a conformal transformation on the four-dimensional metric to 
the Einstein frame
\Eq{
\bar{q}_{\mu\nu}=\left(\frac{\tau\kappa}{\bar{\kappa}}\right)^2q_{\mu\nu}\,,
  \label{ap:ct:Eq}
}
where $\kappa^2$ is the four-dimensional constant and we have chosen 
a volume modulus $\rho$ such that the metric 
$u_{ij}(\Ysp)$ of the internal space is normalized 
$\int d^6y\sqrt{u}=1$\,. 
After performing the conformal transformation, the kinetic terms of 
moduli are given by diagonal form. 
Since these do not have canonical kinetic 
energies, we redefine them \cite{Hertzberg:2007wc}
\Eq{
\bar{\rho}=\sqrt{\frac{3}{2}}\kappa^{-1}\ln \rho\,,~~~~~
\bar{\tau}=\sqrt{{2}}\kappa^{-1}\ln \tau\,.
}
In terms of fields $\bar{\rho}$\,, $\bar{\tau}$\,, the gravity sector in 
the four-dimensional effective action is given by 
\Eqr{
S_{\rm E1}&=&
\int d^4x\sqrt{-\bar{q}}\left[\frac{1}{2\kappa^2}\bar{R}
-\frac{1}{2}\bar{q}^{\mu\nu}\pd_\mu\bar{\rho}\pd_\nu\bar{\rho}
-\frac{1}{2}\bar{q}^{\mu\nu}\pd_\mu\bar{\tau}\pd_\nu\bar{\tau}
\right.\nn\\
&&~~~~~
\left.
+A_\Ysp
\exp\left\{-\kappa\left(\sqrt{2}\bar{\tau}
+\frac{\sqrt{6}}{3}\bar{\rho}\right)\right\}R(\Ysp)\right],
      \label{ap:E1:Eq}
}
where the Ricci scalar $\bar{R}$\,, determinant $\bar{q}$ are defined 
with respect to the metric $\bar{q}_{\mu\nu}$\,,  
$R(\Ysp)$ denotes the Ricci scalar constructed from the metric 
$u_{ij}(\Ysp)$\,, and the coefficient $A_\Ysp$ is the function 
of internal space moduli. 

Next we consider contributions of field strengths  
in the ten-dimensional action 
(\ref{com:action:Eq}). We assume that field strengths $H$ and $F_p$ 
can have a non-vanishing integral over any closed three-, 
$p$-dimensional homology cycle $\Sigma$\,, 
${\cal C}_p$ of the compact space Y, respectively. 
Since these field strengths satisfy the generalized 
Dirac charge quantization condition (\ref{co:Dirac:Eq}), 
the four-dimensional effective potential includes the appropriate 
factors of the volume and dilaton from the compactification. 
The contributions from the field strengths in the four-dimensional 
Einstein frame action can be expressed as 
\Eqr{
S_{\rm E2}
&=&-\int d^4x\sqrt{-\bar{q}}
\left[A_H\exp\left\{-\kappa\left(\sqrt{2}\bar{\tau}
+\sqrt{6}\bar{\rho}\right)\right\}\right.\nn\\
&&\left. +\sum_pA_p
\exp\left[-\kappa\left\{2\sqrt{2}\bar{\tau}
+\frac{\sqrt{6}}{3}(p-3)\bar{\rho}\right\}\right]
\right],
   \label{ap:E2:Eq}
}
where coefficients $A_H$ and $A_p$ depend on the moduli of the 
internal space. 

Finally we consider the four-dimensional effective action coming 
from D$p$-branes, O$p$-planes term in (\ref{com:action:Eq}).  
O$p$-plane is compactified by $(p-3)$-dimensional internal space 
because they extend our four-dimensional universe. 
Then, the four-dimensional Einstein frame action arises from the dimensional 
reduction of the terms in (\ref{com:action:Eq}) associated with 
the various D$p$-branes, O$p$-planes becomes 
\Eqr{
S_{\rm E3}&=&-\int d^4x\sqrt{-\bar{q}}\,
\sum_p\left(A_{{\rm D}p}-A_{{\rm O}p}\right)
\exp\left[-\kappa\left\{\frac{3\sqrt{2}}{2}\bar{\tau}
+\frac{\sqrt{6}}{6}(6-p)\bar{\rho}\right\}\right]\nn\\
&&\times \int d^{p-3}x\sqrt{g_{p-3}}\,.
   \label{ap:E3:Eq}
}
From Eqs.~(\ref{ap:E1:Eq}), (\ref{ap:E2:Eq}) and (\ref{ap:E3:Eq}),  
the four-dimensional effective action $S_{\rm E}$ 
in the Einstein frame is described as 
\Eqr{
S_{\rm E}&=&S_{\rm E1}+S_{\rm E2}+S_{\rm E3}
\nn\\
&=&\int d^4x\sqrt{-\bar{q}}\left[\frac{1}{2\kappa^2}\bar{R}
-\frac{1}{2}\bar{q}^{\mu\nu}\pd_\mu\bar{\rho}\,\pd_\nu\bar{\rho}
-\frac{1}{2}\bar{q}^{\mu\nu}\pd_\mu\bar{\tau}\,\pd_\nu\bar{\tau}
-V\left(\bar{\tau}, \bar{\rho}\right)\right],
}
where $V\left(\bar{\tau}, \bar{\rho}\right)$ is 
the scalar potential \cite{Hertzberg:2007wc}:
\Eqr{
V\left(\bar{\tau}, \bar{\rho}\right)&=&
-A_\Ysp\exp\left[-\kappa\left(\sqrt{2}\bar{\tau}
+\frac{\sqrt{6}}{3}\bar{\rho}\right)\right]R(\Ysp)
+A_H\exp\left[-\kappa\left(\sqrt{2}\bar{\tau}
+\sqrt{6}\bar{\rho}\right)\right]
\nn\\
&&\hspace{-1cm}
+\sum_p A_p
\exp\left[-\kappa\left\{2\sqrt{2}\bar{\tau}
+\frac{\sqrt{6}}{3}(p-3)\bar{\rho}\right\}\right]
\nn\\
&&\hspace{-1cm}+\sum_p\left(A_{{\rm D}p}-A_{{\rm O}p}\right)
\exp\left[-\kappa\left\{\frac{3\sqrt{2}}{2}\bar{\tau}
+\frac{\sqrt{6}}{6}(p-6)\bar{\rho}\right\}\right]
\int d^{p-3}x\sqrt{g_{p-3}}\,.
\label{A:mp:Eq}
}




\begin{thebibliography}{99}

\bibitem{Khoury:2001wf}
  J.~Khoury, B.~A.~Ovrut, P.~J.~Steinhardt and N.~Turok,
  ``The Ekpyrotic universe: Colliding branes and the origin of 
    the hot big bang,''
  Phys.\ Rev.\ D {\bf 64} (2001) 123522
  [hep-th/0103239].

\bibitem{Steinhardt:2001vw}
  P.~J.~Steinhardt and N.~Turok,
  ``A Cyclic model of the universe,''
  Science {\bf 296} (2002) 1436
  [hep-th/0111030].

\bibitem{Erickson:2006wc}
  J.~K.~Erickson, S.~Gratton, P.~J.~Steinhardt and N.~Turok,
  ``Cosmic perturbations through the cyclic ages,''
  Phys.\ Rev.\ D {\bf 75} (2007) 123507
  [hep-th/0607164].

\bibitem{Lehners:2008vx}
  J.~L.~Lehners,
  ``Ekpyrotic and Cyclic Cosmology,''
  Phys.\ Rept.\  {\bf 465} (2008) 223
  [arXiv:0806.1245 [astro-ph]].

\bibitem{Lehners:2009eg}
  J.~L.~Lehners, P.~J.~Steinhardt and N.~Turok,
  ``The Return of the Phoenix Universe,''
  Int.\ J.\ Mod.\ Phys.\ D {\bf 18} (2009) 2231
  [arXiv:0910.0834 [hep-th]].

\bibitem{Khoury:2009my}
  J.~Khoury and P.~J.~Steinhardt,
  ``Adiabatic Ekpyrosis: Scale-Invariant Curvature Perturbations 
    from a Single Scalar Field in a Contracting Universe,''
  Phys.\ Rev.\ Lett.\  {\bf 104} (2010) 091301
  [arXiv:0910.2230 [hep-th]].

\bibitem{Lyth:2001pf}
  D.~H.~Lyth,
  ``The Primordial curvature perturbation in the ekpyrotic universe,''
  Phys.\ Lett.\ B {\bf 524} (2002) 1
  [hep-ph/0106153].

\bibitem{Brandenberger:2001bs}
  R.~Brandenberger and F.~Finelli,
  ``On the spectrum of fluctuations in an effective field theory of the Ekpyrotic universe,''
  JHEP {\bf 0111} (2001) 056
  [hep-th/0109004].

\bibitem{Hwang:2001ga}
  J.~C.~Hwang,
  ``Cosmological structure problem in the ekpyrotic scenario,''
  Phys.\ Rev.\ D {\bf 65} (2002) 063514
  [astro-ph/0109045].

\bibitem{Khoury:2001zk}
  J.~Khoury, B.~A.~Ovrut, P.~J.~Steinhardt and N.~Turok,
  ``Density perturbations in the ekpyrotic scenario,''
  Phys.\ Rev.\ D {\bf 66} (2002) 046005
  [hep-th/0109050].

\bibitem{Lyth:2001nv}
  D.~H.~Lyth,
  ``The Failure of cosmological perturbation theory 
    in the new ekpyrotic scenario,''
  Phys.\ Lett.\ B {\bf 526} (2002) 173
  [hep-ph/0110007].

\bibitem{Tsujikawa:2001ad}
  S.~Tsujikawa,
  ``Density perturbations in the ekpyrotic universe and 
    string inspired generalizations,''
  Phys.\ Lett.\ B {\bf 526} (2002) 179
  [gr-qc/0110124].

\bibitem{Notari:2002yc}
  A.~Notari and A.~Riotto,
  ``Isocurvature perturbations in the ekpyrotic universe,''
  Nucl.\ Phys.\ B {\bf 644} (2002) 371
  [hep-th/0205019].

\bibitem{Tsujikawa:2002qc}
  S.~Tsujikawa, R.~Brandenberger and F.~Finelli,
  ``On the construction of nonsingular pre - big bang 
    and ekpyrotic cosmologies and the resulting density perturbations,''
  Phys.\ Rev.\ D {\bf 66} (2002) 083513
  [hep-th/0207228].

\bibitem{Lehners:2007ac}
  J.~L.~Lehners, P.~McFadden, N.~Turok and P.~J.~Steinhardt,
  ``Generating ekpyrotic curvature perturbations before the big bang,''
  Phys.\ Rev.\ D {\bf 76} (2007) 103501
  [hep-th/0702153 [hep-th]].

\bibitem{Buchbinder:2007ad}
  E.~I.~Buchbinder, J.~Khoury and B.~A.~Ovrut,
  ``New Ekpyrotic cosmology,''
  Phys.\ Rev.\ D {\bf 76} (2007) 123503
  [hep-th/0702154].

\bibitem{Koyama:2007ag}
  K.~Koyama, S.~Mizuno and D.~Wands,
  ``Curvature perturbations from ekpyrotic collapse with multiple fields,''
  Class.\ Quant.\ Grav.\  {\bf 24} (2007) 3919
  [arXiv:0704.1152 [hep-th]].

\bibitem{Koyama:2007if}
  K.~Koyama, S.~Mizuno, F.~Vernizzi and D.~Wands,
  ``Non-Gaussianities from ekpyrotic collapse with multiple fields,''
  JCAP {\bf 0711} (2007) 024
  [arXiv:0708.4321 [hep-th]].

\bibitem{Buchbinder:2007at}
  E.~I.~Buchbinder, J.~Khoury and B.~A.~Ovrut,
  ``Non-Gaussianities in new ekpyrotic cosmology,''
  Phys.\ Rev.\ Lett.\  {\bf 100} (2008) 171302
  [arXiv:0710.5172 [hep-th]].

\bibitem{Lehners:2007wc}
  J.~L.~Lehners and P.~J.~Steinhardt,
  ``Non-Gaussian density fluctuations from entropically generated curvature perturbations in Ekpyrotic models,''
  Phys.\ Rev.\ D {\bf 77} (2008) 063533
   Erratum: [Phys.\ Rev.\ D {\bf 79} (2009) 129903]
  [arXiv:0712.3779 [hep-th]].

\bibitem{Lehners:2008my}
  J.~L.~Lehners and P.~J.~Steinhardt,
  ``Intuitive understanding of non-gaussianity in ekpyrotic 
    and cyclic models,''
  Phys.\ Rev.\ D {\bf 78} (2008) 023506
   Erratum: [Phys.\ Rev.\ D {\bf 79} (2009) 129902]

\bibitem{Mizuno:2008zza}
  S.~Mizuno, K.~Koyama, F.~Vernizzi and D.~Wands,
  ``Primordial non-Gaussianities in new ekpyrotic cosmology,''
  AIP Conf.\ Proc.\  {\bf 1040} (2008) 121.

\bibitem{Linde:2009mc}
  A.~Linde, V.~Mukhanov and A.~Vikman,
  ``On adiabatic perturbations in the ekpyrotic scenario,''
  JCAP {\bf 1002} (2010) 006
  [arXiv:0912.0944 [hep-th]].

\bibitem{Lehners:2010fy}
  J.~L.~Lehners,
  ``Ekpyrotic Non-Gaussianity: A Review,''
  Adv.\ Astron.\  {\bf 2010} (2010) 903907
  [arXiv:1001.3125 [hep-th]].

\bibitem{Li:2013hga}
  M.~Li,
  ``Note on the production of scale-invariant entropy perturbation in the Ekpyrotic universe,''
  Phys.\ Lett.\ B {\bf 724} (2013) 192
  [arXiv:1306.0191 [hep-th]].

\bibitem{Battarra:2013cha}
  L.~Battarra and J.~L.~Lehners,
  ``Quantum-to-classical transition for ekpyrotic perturbations,''
  Phys.\ Rev.\ D {\bf 89} (2014) no.6,  063516
  [arXiv:1309.2281 [hep-th]].

\bibitem{Fertig:2013kwa}
  A.~Fertig, J.~L.~Lehners and E.~Mallwitz,
  ``Ekpyrotic Perturbations With Small Non-Gaussian Corrections,''
  Phys.\ Rev.\ D {\bf 89} (2014) no.10,  103537
  [arXiv:1310.8133 [hep-th]].

\bibitem{Ijjas:2014fja}
  A.~Ijjas, J.~L.~Lehners and P.~J.~Steinhardt,
  ``General mechanism for producing scale-invariant 
    perturbations and small non-Gaussianity in ekpyrotic models,''
  Phys.\ Rev.\ D {\bf 89} (2014) no.12,  123520
  [arXiv:1404.1265 [astro-ph.CO]].

\bibitem{Levy:2015awa}
  A.~M.~Levy, A.~Ijjas and P.~J.~Steinhardt,
  ``Scale-invariant perturbations in ekpyrotic cosmologies 
   without fine-tuning of initial conditions,''
  Phys.\ Rev.\ D {\bf 92} (2015) no.6,  063524
  [arXiv:1506.01011 [astro-ph.CO]].

\bibitem{Fertig:2015ola}
  A.~Fertig and J.~L.~Lehners,
  ``The Non-Minimal Ekpyrotic Trispectrum,''
  JCAP {\bf 1601} (2016) no.01,  026
  [arXiv:1510.03439 [hep-th]].

\bibitem{Ito:2016fqp}
  A.~Ito and J.~Soda,
  ``Primordial Gravitational Waves Induced by Magnetic Fields 
   in an Ekpyrotic Scenario,''
  Phys.\ Lett.\ B {\bf 771} (2017) 415
  [arXiv:1607.07062 [hep-th]].

\bibitem{Gratton:2003pe}
  S.~Gratton, J.~Khoury, P.~J.~Steinhardt and N.~Turok,
  ``Conditions for generating scale-invariant density perturbations,''
  Phys.\ Rev.\ D {\bf 69} (2004) 103505
  [astro-ph/0301395].

\bibitem{Khoury:2003rt}
  J.~Khoury, P.~J.~Steinhardt and N.~Turok,
  Designing cyclic universe models,''
  Phys.\ Rev.\ Lett.\  {\bf 92} (2004) 031302
  [hep-th/0307132].

\bibitem{Hertzberg:2007wc}
  M.~P.~Hertzberg, S.~Kachru, W.~Taylor and M.~Tegmark,
  ``Inflationary Constraints on Type IIA String Theory,''
  JHEP {\bf 0712} (2007) 095
  [arXiv:0711.2512 [hep-th]].

\bibitem{Haque:2008jz}
  S.~S.~Haque, G.~Shiu, B.~Underwood and T.~Van Riet,
  ``Minimal simple de Sitter solutions,''
  Phys.\ Rev.\ D {\bf 79} (2009) 086005
  [arXiv:0810.5328 [hep-th]].

\bibitem{Giddings:2001yu}
  S.~B.~Giddings, S.~Kachru and J.~Polchinski,
  ``Hierarchies from fluxes in string compactifications,''
  Phys.\ Rev.\ D {\bf 66} (2002) 106006
  [hep-th/0105097].

\bibitem{Kachru:2003aw}
  S.~Kachru, R.~Kallosh, A.~D.~Linde and S.~P.~Trivedi,
  ``De Sitter vacua in string theory,''
  Phys.\ Rev.\ D {\bf 68} (2003) 046005
  [hep-th/0301240].

\bibitem{Villadoro:2005cu}
  G.~Villadoro and F.~Zwirner,
  ``$N=1$ effective potential from dual type-IIA D6/O6 
    orientifolds with general fluxes,''
  JHEP {\bf 0506} (2005) 047
  [hep-th/0503169].

\bibitem{DeWolfe:2005uu}
  O.~DeWolfe, A.~Giryavets, S.~Kachru and W.~Taylor,
  ``Type IIA moduli stabilization,''
  JHEP {\bf 0507} (2005) 066
  [hep-th/0505160].

\bibitem{deCarlos:2009fq}
  B.~de Carlos, A.~Guarino and J.~M.~Moreno,
  ``Flux moduli stabilisation, Supergravity algebras and no-go theorems,''
  JHEP {\bf 1001} (2010) 012
  [arXiv:0907.5580 [hep-th]].

\bibitem{Silverstein:2007ac}
  E.~Silverstein,
  ``Simple de Sitter Solutions,''
  Phys.\ Rev.\ D {\bf 77} (2008) 106006
  [arXiv:0712.1196 [hep-th]].

\bibitem{Danielsson:2009ff}
  U.~H.~Danielsson, S.~S.~Haque, G.~Shiu and T.~Van Riet,
  ``Towards Classical de Sitter Solutions in String Theory,''
  JHEP {\bf 0909} (2009) 114
  [arXiv:0907.2041 [hep-th]].

\bibitem{Meeus:2016}
Eline Meeus, ``A nogo-theorem for Ekpyrosis from 10D supergravity,'' 
master's thesis, KU Leuven (2016).

\bibitem{Gibbons:1984kp}
  G.~W.~Gibbons,
  ``Aspects Of Supergravity Theories,''
  Print-85-0061 (CAMBRIDGE).

\bibitem{deWit:1986mwo}
  B.~de Wit, D.~J.~Smit and N.~D.~Hari Dass,
  ``Residual Supersymmetry of Compactified $D=10$ Supergravity,''
  Nucl.\ Phys.\ B {\bf 283} (1987) 165.

\bibitem{Maldacena:2000mw}
  J.~M.~Maldacena and C.~Nunez,
  ``Supergravity description of field theories on curved manifolds 
    and a no go theorem,''
  Int.\ J.\ Mod.\ Phys.\ A {\bf 16} (2001) 822
  [hep-th/0007018].

\bibitem{Caviezel:2009tu}
  C.~Caviezel, T.~Wrase and M.~Zagermann,
  ``Moduli Stabilization and Cosmology of Type IIB on SU(2)-Structure 
    Orientifolds,''
  JHEP {\bf 1004} (2010) 011
  [arXiv:0912.3287 [hep-th]].

\bibitem{Wrase:2010ew}
  T.~Wrase and M.~Zagermann,
  ``On Classical de Sitter Vacua in String Theory,''
  Fortsch.\ Phys.\  {\bf 58} (2010) 906
  [arXiv:1003.0029 [hep-th]].

\bibitem{Shiu:2011zt}
  G.~Shiu and Y.~Sumitomo,
  ``Stability Constraints on Classical de Sitter Vacua,''
  JHEP {\bf 1109} (2011) 052
  [arXiv:1107.2925 [hep-th]].

\bibitem{VanRiet:2011yc}
  T.~Van Riet,
  ``On classical de Sitter solutions in higher dimensions,''
  Class.\ Quant.\ Grav.\  {\bf 29} (2012) 055001
  [arXiv:1111.3154 [hep-th]].

\end{thebibliography}
\end{document}